\documentclass[preprintnumbers, floatfix, preprintnumbers, letterpaper, twocolumn, superscriptaddress,nofootinbib]{revtex4}
\usepackage{graphicx}
\usepackage{microtype}
\usepackage{amsmath}
\usepackage{amssymb}
\usepackage{subfigure}
\usepackage{hyperref}
\usepackage{url}
\usepackage{xcolor}
\usepackage{color}
\usepackage{mathrsfs}
\usepackage{calrsfs}
\usepackage{amsfonts}
\usepackage{tabularx}
\usepackage{latexsym}
\usepackage{ragged2e}
\usepackage{epsfig}
\usepackage{textcomp}
\usepackage{float}

\usepackage{caption}
\DeclareCaptionJustification{justified}{\leftskip=0pt \rightskip=0pt \parfillskip=0pt plus 1fil}
\definecolor{vividviolet}{rgb}{0.62, 0.0, 1.0}
\definecolor{amaranth}{rgb}{0.9, 0.17, 0.31}
\definecolor{palatinateblue}{rgb}{0.15, 0.23, 0.89}
\definecolor{brightpink}{rgb}{1.0, 0.0, 0.5}
\definecolor{cornflowerblue}{rgb}{0.39, 0.58, 0.93}
\definecolor{deepcarminepink}{rgb}{0.94, 0.19, 0.22}
\definecolor{radicalred}{rgb}{1.0, 0.21, 0.37}

\hypersetup{ linktoc=all,
	colorlinks, linkcolor={palatinateblue},
	citecolor={brightpink}, urlcolor={amaranth}
}

\graphicspath{{Images/}}

\renewcommand{\d}[1]{\ensuremath{\operatorname{d}\!{#1}}}

\def\sideremark#1{\ifvmode\leavevmode\fi\vadjust{\vbox to0pt{\vss
			\hbox to 0pt{\hskip\hsize\hskip1em
				\vbox{\hsize1.3cm\tiny\raggedright\pretolerance10000
					\noindent #1\hfill}\hss}\vbox to8pt{\vfil}\vss}}}%
\def\beq{\begin{equation}}
\def\eeq{\end{equation}}

\begin{document}
\title{A Maximum Force Perspective on Black Hole Thermodynamics, \\Quantum Pressure, and Near-Extremality}

\author{Yen Chin \surname{Ong}}
\email{ycong@yzu.edu.cn}
\affiliation{Center for Gravitation and Cosmology, College of Physical Science and Technology, Yangzhou University, \\180 Siwangting Road, Yangzhou City, Jiangsu Province  225002, China}
\affiliation{Shanghai Frontier Science Center for Gravitational Wave Detection, School of Aeronautics and Astronautics, Shanghai Jiao Tong University, Shanghai 200240, China}

\begin{abstract}
I re-examined the notion of the thermodynamic force constructed from the first law of black hole thermodynamics. In general relativity, the value of the charge (or angular momentum) at which the thermodynamic force equals the conjectured maximum force $F=1/4$ is found to correspond to $Q^2/M^2=8/9$ (respectively, $a^2/M^2=8/9$), which is known in the literature to exhibit some special properties. This provides a possible characterization of near-extremality. In addition, taking the maximum force conjecture seriously amounts to introducing a pressure term in the first law of black hole thermodynamics. This resolves the factor of two problem between the proposed maximum value $F=1/4$ and the thermodynamic force of Schwarzschild spacetime $F=1/2$. Surprisingly it also provides another indication for the instability of the inner horizon. For a Schwarzschild black hole, under some reasonable assumptions, this pressure can be interpreted as being induced by the quantum fluctuation of the horizon position, effectively giving rise to a diffused ``shell'' of characteristic width $\sqrt{M}$. The maximum force can therefore, in some contexts, be associated with inherently quantum phenomena, despite the fact that it is free of $\hbar$. Some implications are discussed as more questions are raised.
\end{abstract}

\maketitle
\section{Introduction: Maximum Force and The Mystery of a Factor of Two}
In \cite{1809.00442}, I noted that the first law of thermodynamics for a Schwarzschild black hole $\d M=T\d S$, where $M$, $T$ and $S$ are the black hole mass, Hawking temperature and the Bekenstein-Hawking entropy respectively, can be expressed as a ``thermodynamic force''\footnote{In the following I will mostly set $c=G=\hbar=k_B=1$, but will restore them when emphasis is needed.}
\begin{equation}\label{1}
F_{\text{therm}}:=\frac{\d M}{\d r_+}=T\frac{\d S}{\d r_+}=\frac{c^4}{2G},
\end{equation}
where $r_+$ denotes the horizon.
On the other hand, in general relativity (GR), the ``maximum force conjecture'' states that there exists an upper bound for forces acting between two bodies \cite{0210109,1408.1820,0607090,724159}: $F \leqslant F_{\text{max}}=1/4$. 
This conjecture has attracted quite some attentions and controversies recently, see, e.g., \cite{1504.01547,2102.01831,2105.07929,2112.15418,2109.07700,PhysRevD.104.068502,2207.02465,2108.13435}. 

In this work, I will only discuss the thermodynamic force of (asymptotically flat 4-dimensional) black holes in GR; I am less concerned about the validity of the conjectures in more general contexts.
It is worth emphasizing that while we often do not make use of ``forces'' in GR, there is no issue in defining\footnote{One objection is that this definition seems to depend on the coordinate choice. We defer this to the Discussion.} Eq.(\ref{1}). 
Since $F_{\text{therm}}>F_{\text{max}}$, I proposed in \cite{1809.00442} the ``weak maximum force conjecture'' (WMFC), in which $F_{\text{max}}=O(1)$, to distinguish it from the original strong form of the conjecture (SMFC). For Schwarzschild black holes, $F_{\text{therm}}=1/2=2F_{\text{max}}$, which I will refer herein as the ``factor of two problem''. 

In \cite{2112.15418}, Schiller (see also \cite{2205.06302}) objected that previous works \cite{1809.00442,9912110,2108.07407} with the maximum force value being 1/2 instead of 1/4 failed to take into account either the difference between radius and diameter, or the factor of 2 in the Smarr relation for black holes ($M=2TS$). However, it can readily be checked that Eq.(\ref{1}) is correct and in my opinion cannot be explained by either of these reasons (The Smarr relation is definitely correct as Eq.(\ref{1}) is just the first law of black hole thermodynamics; for the issue of diameter vs radius, we defer to the Discussion section). In fact, the argument in \cite{2112.15418} itself amounts to the incorrect relation $M=TS$, not $M=2TS$. To see this\footnote{Or note that Eq.(11) in Ref.(\cite{2112.15418}) is the first law of black hole thermodynamics $\d M=T \d S$, so its Eq.(10) is equivalent to $M=TS$.}, we note that Eq.(10) in Ref.(\cite{2112.15418}) gives 
\begin{equation}\label{e2} E =\frac{\kappa}{8 \pi} A = \frac{\kappa}{2 \pi} S = \frac{1}{8 \pi M} S = TS. \end{equation} 
This result was obtained by heuristically considering, assuming $F=1/4$ holds, the equalities
\begin{equation}\label{e3}
\frac{F}{A} = \frac{1/4}{4\pi r_+^2} = \frac{E/L}{A},
\end{equation}
where $E=M$ is the energy of the system, and the size of the system $L \leqslant r_+ = {1}/{2\kappa}$. Putting aside the issue whether $r_+$ can be taken as a measure of the system ``size''\footnote{One might object that in general relativity, due to spacetime curvature the Schwarzschild $r$ coordinate is only an areal radius, not a proper length. Nevertheless, there is a good reason to take $2M$ as the size of the black hole -- see the Discussion section.}, it would seem that the SMFC value $F=1/4$ is inconsistent with $M=2TS$. This is another manifestation of the factor of two problem. $M=2TS$ can be obtained if we use $F=1/2$ instead. However, as we shall see the issue is deeper than this. To this end, we need to consider more general black holes, namely how $F$ changes if rotation is included, for example.

In this note, I would like to point out -- even if black holes only conform to WMFC, this does not mean that the value $F=1/4$ has no significance in this context. 
To this end, I shall first discuss near-extremal black holes and how they led to the consideration of SMFC in Sec.(\ref{II}). In doing so we will also see in Sec.(\ref{IIIa}) that as a bonus, SMFC is consistent with the well-known instability of the inner horizon. 
In Sec.(\ref{IIIb}), we will see that this may point towards the possibility that the first law we considered is incomplete, and that \emph{SMFC may in fact, hold}, but for nontrivial reasons. This would resolve the factor of two issue and reconcile both the WMFC and SMFC, at least in the contexts of black hole thermodynamics. More specifically, as we will see, WMFC holds if we treat black holes ``classically'', in the sense that the $T \d S$ term in the first law is interpreted as $(\kappa/8\pi)\d A$ as it was originally conceived \cite{BCH}; i.e. $\hbar$ cancels out from the numerator of $T$ and the denominator of $S$. On the other hand, SMFC holds if we allow quantum fluctuation on the horizon position, which induces a pressure term. In other words, SMFC in this context is \emph{inherently quantum} in nature. 
This is clarified in Sec.(\ref{IV}). This may seem odd to readers who are familiar with the maximum force conjecture since it is often emphasized that $F$ is a classical quantity independent of $\hbar$ (see, e.g. \cite{2006.07338}). I will address this in the Discussion section (Sec.(\ref{V})). . 

\section{Interlude: How Close to Extremal Is Near-Extremal?}\label{II}

The mass function $M(Q,r_+):=(Q^2+r_+^2)/2r_+$ for the Reissner-Nordstr\"om black hole (analogously for the Kerr case) leads to the definition of the thermodynamic force
\begin{equation}
F_{\text{therm}}:=\frac{\partial M}{\partial r_+}=\frac{\sqrt{1-x^2}}{1+\sqrt{1-x^2}}, 
\end{equation}
where $x:=Q/M$ (or $a/M$ for the Kerr case; without loss of generality, let us assume $Q,a>0$). 
This expression is monotonically decreasing from $1/2$ to $0$ in the extremal limit. Remarkably, the value of $x$ at which $F$ attains the purported SMFC value, namely $F=1/4$, corresponds to $x=\sqrt{8/9}\approx 0.9428$. This is a value that corresponds to some interesting phenomena already noticed by various authors in the literature. 

For Reissner-Nordstr\"om black holes with $x > \sqrt{8/9}$, the effective Hawking temperature at the horizon is negative, as recently shown in \cite{2301.12319} by McMaken and Hamilton. 
{Unlike the Hawking temperature $T=\kappa/2\pi$, which is observed by an asymptotic observer, this effective temperature is obtained from an effective ``surface gravity'' defined as
\begin{equation}
\kappa(u):=\frac{d}{du} \ln\left({\frac{dU}{du}}\right),
\end{equation}
where $u$ denotes the outgoing null coordinate of the observer's position and $U$ denotes the position of an emitter that defines the vacuum state. One can similarly define an effective temperature for the inner horizon. As mentioned in \cite{2301.12319}, as long as $\kappa(u)$ is approximately constant over a small interval around some $u^*$, the vacuum expectation value of the particle number operator is consistent with
that of a Planckian spectrum with temperature $\kappa(u^*)/2\pi$.
Operationally it is often more convenient to work with
\begin{equation}
\kappa(u):=-\frac{d}{d\tau_\text{ob}} \ln\left({\frac{\omega_\text{ob}}{\omega_\text{em}}}\right),
\end{equation}
where $\tau_\text{ob}$ is the proper time of the observer, while
$\omega_\text{ob}$ and $\omega_\text{em}$ are the frequencies measured in the frame of the observer and emitter, respectively. The statement that the effective Hawking temperature at the horizon is negative for $x > \sqrt{8/9}$ specifically refers to the temperature as seen by an observer in free fall from rest at infinity towards the black hole, with the emitter located at the horizon.}

In my earlier work with Good \cite{2003.10429}, we showed essentially the same result heuristically using a gravitational analog of Schwinger effect\footnote{Our argument was based on the radial tidal force, which for the Reissner-Nordstr\"om spacetime, is given by $a^r=\left(2M/r^3-3Q^2/r^4\right)n^r$ \cite{1602.07232}. Let $r_*$ denotes the position at which $a^r$ changes sign. It can be shown that $r_*=r_+$ precisely when $x = \sqrt{8/9}$.}, which revealed that the frequency of a typical Hawking particle becomes negative near the horizon for $x > \sqrt{8/9}$. 
Similarly, by utilizing an embedding method Brynjolfsson
and Thorlacius \cite{0805.1876} argued that a freely falling observer would not detect any radiation near the black hole when $x > \sqrt{8/9}$. This peculiarity was also reflected in the stress-energy tensor expectation in the $(1+1)$-dimensional analysis of Loranz and Hiscock \cite{9607048}, for which $\langle T_t^{~t}(r) \rangle$ changes sign at $x=\sqrt{8/9}$.
Naturally one would ask what is the physical interpretation for a negative temperature. In \cite{2301.12319} we see that the inner horizon has an infinite negative effective temperature, which could be indicative of its unstable nature. Whether a negative yet finite temperature is associated to any peculiarity in the particle production or spectrum remains to be further studied \cite{2305.09019}. In some models, e.g., \cite{1404.0602,1405.5235}, a negative energy flux is emitted during the Hawking process, which causes the black hole to temporarily increase its mass during evaporation. In \cite{1506.08072}, Good and I studied a moving mirror model that reproduces a similar behavior but found no sign of anything peculiar in the particle emission. Likewise, the effective negative temperature discussed here may give no particle production, as also mentioned in \cite{0805.1876}. However, the point is that, the result that Reissner-Nordstr\"om black holes behave differently in the regime $(Q/M)^2\geqslant 8/9$ seems to be quite robust, having been obtained via quite varied approaches. (We also note that the stability of Reissner-Nordstr\"om black holes against charged scalar perturbations were proved separately for the regime $(Q/M)^2\leqslant 8/9$ and $8/9 <(Q/M)^2\ <1$. See \cite{1304.6474} and \cite{1504.00009} respectively.)

For Kerr black holes\footnote{There are other instances at which this value turns up, but they are most probably irrelevant to our discussions. For example, in 5-dimensions the entropy of a
``large black ring'' (with unbounded angular momentum) exceeds that of the singly-rotating black hole of the same mass at $x=\sqrt{8/9}$ (here $x$ is the 5-dimensional version of the dimensionless rotation parameter) \cite{0110260}. The value also occurs in the context of gravitational entropy of Kerr black holes \cite{1407.6941}. Also, in astrophysics, the maximum spin-equilibrium accretion efficiency occurs at $x \approx 0.94$ for the thin disk model considered in \cite{9908049}.}, I am not aware of
any change in the effective Hawking temperature at $x=\sqrt{8/9}$, though recently  Dai and Stojkovic found that Hawking emission becomes sub-dominant compared to superradiant radiation precisely near $x\approx 0.94$ \cite{2306.17423}. 
At the classical level, it is well known that the equatorial innermost stable circular orbit (ISCO) lies inside the ergosphere when $x > \sqrt{8/9}$, which can be checked from the equations given in \cite{bardeen}; see also the Appendix of \cite{0911.3889}. Such rapidly spinning black holes play important roles in astrophysics \cite{1204.5854,2202.06958}. From the quantum prespective, a rapidly spinning Kerr black hole does display a distinctly different behavior compared to slowly rotating one: its emission spectra would eventually become continuous near extremality $x \gtrsim 0.9$ \cite{1909.04057}, though the exact value of the discrete/continuous transition characterized by the ratio $\tau/\gamma T$ (where $\tau$ is the characteristic black hole lifetime under Hawking evaporation and $\gamma$ is a dimensionless constant of order unity that specifies the discrete spectrum of the quantized horizon) depends on the exact choice of $\gamma$. It might be interesting to check if setting $\tau/\gamma T=1$ at $x=\sqrt{8/9}$ would give us a reasonable value of $\gamma$ that is supported by other arguments.

Note that in the mass function, $Q$ and $a$ are treated as the black hole parameters independent of $r_+$. Therefore $F_{\text{therm}}$ for Reissner-Nordstr\"om black holes is equal to
\begin{equation}
F_{\text{therm}}=T\frac{\partial S}{\partial r_+} + \Phi_+ \underbrace{\frac{\partial Q}{\partial r_+}}_{=0}=T\frac{\partial S}{\partial r_+}=2\pi T r_+ ,
\end{equation}
where $\Phi_+:=Q/r_+$ is the electric potential at the horizon.
For the Kerr case, on the other hand,
\begin{equation}
F_{\text{therm}}=T\frac{\partial S}{\partial r_+} + \Omega_+ \frac{\partial J}{\partial r_+}.
\end{equation}
Unlike the charged case, the second term on the right hand side contributes since the angular momentum is $J=aM$ and $M$ is given by $M=(a^2+r_+^2)/2r_+$. However, note that conveniently, $F_{\text{therm}}$ can be computed directly from the differentiation of the mass with respect to the horizon radius, without working through the first law.

I should emphasize that up until now these are just standard black hole thermodynamics, re-expressed in terms of the thermodynamic force. We thus see that the SMFC value for the thermodynamic force $F_\text{therm}=1/4$ picks up quite special value of the charge-to-mass or rotation parameter-to-mass ratio of the black holes.

As already mentioned, McMaken and Hamilton \cite{2301.12319} showed that the \emph{inner horizon} of a Reissner-Nordstr\"om black hole is associated with a negative temperature (see also \cite{OK, 1206.2015, 1806.11134, 2107.11193, liu}). They further proposed that when $Q^2/M^2 > 8/9$, the fact that the effective Hawking temperature near the event horizon becomes negative is due to the inner horizon being ``close enough'' to the event horizon that the negative temperature can be detected \emph{outside} the black hole. If this picture is correct, $x^2=8/9$ could mark the transition for when some properties of the interior become prominent. This is extremely interesting in view of recent arguments that quantum effects become important at the horizon of near-extremal black holes \cite{2210.02473,2303.07358} (see also \cite{2307.10423} and some earlier works \cite{1005.2999,1105.2574}). The criterion $x^2 \geqslant 8/9$ can thus be taken as a characterization for what it means to be ``near-extremal'' or ``quantum dominated'', for a 4-dimensional asymptotically flat Reissner-Nordstr\"om spacetime. While $\sqrt{8/9} \approx 0.943$ might seem -- at first impression -- to be still far away from extremality, it is comparable to the charge-to-mass ratio that asymptotically locally anti-de Sitter black holes whose horizon has a torus topology becomes unstable against brane pair production due to stringy effects (said ratio is 0.916 in $\text{AdS}_4$ and 0.958 in $\text{AdS}_5$) \cite{1012.4056} -- i.e., classical GR solution ceases to be a good description of the bulk physics. At least in spirit, this holographic result is similar to the preceding claim that black holes with $x\geqslant \sqrt{8/9}$ should be characterized as ``quantum dominated''. Granted that the results in \cite{2301.12319} do not apply to Kerr black holes, near-extremal Kerr black holes are also highly quantum \cite{2303.07358}, so it is not too far-fetched to suggest that if one takes the maximum force conjecture as a guide, the same criterion should hold.

\section{Maximum Force and the Inner Horizon Instability}\label{IIIa}

If SMFC is indeed correct, then requiring that it holds may reveal some interesting physics. (In fact, if SMFC does not hold, it would be a remarkable coincidence that $F_{\text{therm}}=1/4$ corresponds \emph{exactly} to the near-extremal transition with the aforementioned behaviors found in the literature.) In spirit, this is similar to imposing cosmic censorship in \cite{1907.07490}, which led to the correct nontrivial production rate of charged particles in a dilaton black hole background (which can be derived using QFT independently without any mention of cosmic censorship). 
In view of this prospect, I shall propose that it might be insightful to define a shifted quantity $f:=F_{\text{therm}}+\tilde{F}$, where $\tilde{F}$ is the shifted amount, such that $|f|\leqslant 1/4$ satisfies the SMFC. (Other form of $f=f(F_\text{therm})$ may be possible, but let us keep to the simplest option in this work.)

One obvious choice is to set $\tilde{F}\equiv -1/4$. Then,
in terms of $f$, a Schwarzschild black hole would saturate the original maximum force $f=1/4$, but so would an extremal black hole with $f=-1/4$ though with an opposite sign. In other words, a Schwarzschild black hole and an extremal black hole sit on the boundary of the SMFC bound, while non-extremal holes can take any value $|f|<1/4$.
With this choice of $\tilde{F}$, classical black holes satisfy $f>0$ whereas $f<0$ corresponds to quantum dominated black holes. Though not quite in the same context, it is interesting to note that in the \emph{massless} limit (for fixed $a$), the Kerr geometry can be interpreted as a cosmic string with a negative tension $T=-1/4$ \cite{1705.07787,1606.04879,1701.05533}; see also \cite{9607008}. (Incidentally, Hiscock showed that cosmic strings with tension magnitude greater than $1/4$ would result in the collapse of the exterior geometry \cite{hiscock}, which is also in accordance with SMFC. The relation between the maximum force and cosmic strings was also noticed in \cite{0210109,1408.1820}.)

\begin{figure}[htbp]
\centering
\includegraphics[width=1.00\columnwidth,keepaspectratio]{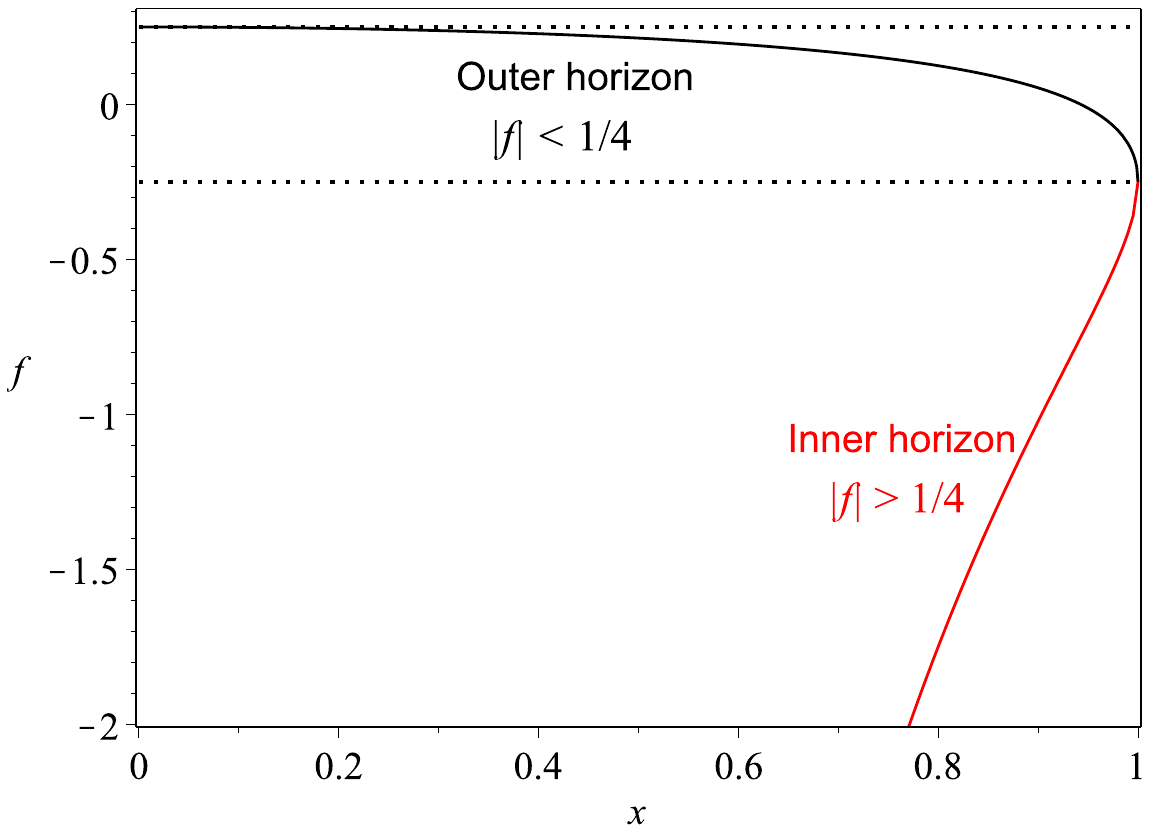}
\caption{The shifted thermodynamic force $f$, as a function of $x$ ($x$ equals $Q/M$ or $x=a/M$ for Reissner-Nordstr\"om and Kerr black holes, respectively), for the outer horizon is always bounded between $-1/4$ and $1/4$ ($f=0$ corresponds to $x= \sqrt{8/9}$), which are indicated by the dotted lines. On the other hand, the inner horizon satisfies $f<-1/4$, which violates the SMFC. In fact, $f$ is not bounded from below for the inner horizon -- it tends to $-\infty$ in the $x\to 0$ limit.
}\label{f}
\end{figure}

Another remarkable implication analogous to the cosmic string collapse follows: the instability of the inner horizon is indicated by a simple computation that its associated shifted force satisfies $f<-1/4$; i.e., $f$ \emph{violates} SMFC. To see this, we note that from the mass function $M(r_-,x)=(x^2+r_-^2)/2r_-$, we can differentiate with respect to $r_-$ to obtain the thermodynamic force of the inner horizon. The obtained expression is
\begin{equation}
F_{\text{therm}}[r_-]=\frac{1-x^2-\sqrt{1-x^2}}{(1-\sqrt{1-x^2})^2}.
\end{equation}
The shifted quantity $f:=F_{\text{therm}}[r_-]-1/4$ is then a monotonic function that goes to $-1/4$ in the extremal limit, and diverges to $-\infty$ in the $x \to 0$ limit.
In other words $f_\text{inner horizon}\in (-\infty, -1/4]$, see Fig.(\ref{f}). Just like Hiscock's result about the collapse of cosmic strings once its tension exceeds the maximum force bound, this is in agreement with the instability of the inner horizon due to mass inflation (blueshift instability) \cite{87999,1978.0024,1663,004,1704.05790,1902.08323,1912.10890,2001.11156,2105.04604}. Furthermore, as argued in \cite{2301.12319} in the context of Reissner-Nordstr\"om spacetime, there is most likely a runaway particle creation at the inner horizon, as the result of which the inner
horizon must collapse into a singularity (or the geometry may evolve dynamically into something else entirely). One may wonder why $f$ violates the maximum force the most in the $x \to 0$ limit. However, this is again consistent with \cite{2301.12319}, which showed that the particle spectrum diverges at \emph{all} frequencies as $Q/M \to 0$, whereas the blueshift is relatively less severe for large $Q/M$. The physical reason for this is that in the $x \to 0$ limit, the distance of the inner horizon to the spacelike singularity is closer \cite{2301.12319}. 

This finding regarding the inner horizon thermodynamic force is also consistent with previous arguments in the literature that the maximum force is related to the cosmic censorship conjecture \cite{1408.1820,2005.0680}. However here it is more relevant to the strong cosmic censorship, which requires the inner horizon to be \emph{unstable}. (Thus to be more specific, it is not that forces cannot exceed $1/4$, but rather that such a violation would lead to instabilities or other pathologies).

\section{Maximum Force Gives Rise to Pressure in the First Law}\label{IIIb}

The definition of $f:=F_{\text{therm}}+\tilde{F}$ means that in addition to the horizon thermodynamics that gives rise to $F_{\text{therm}}$, there is also an extra force contribution. One possibility is that $\tilde{F}$ corresponds to the black hole interior, but need not necessarily be related to the inner horizon $r_-$, which is in any case, unstable, as mentioned above. Or it could be related to the near-horizon region of the exterior spacetime.
In any case, quite naturally, we may interpret this extra force term as the result of a pressure term in the first law.
That is, if we focus on the Schwarzschild case for simplicity, $\d E=T\d S  - P\d V$. 

Now, the question is the following: what is the volume $V$? We need to fix $V$ in order to find $P$. 
I suggest that a sensible choice is 
\begin{equation}
V=4\pi r_+^2\varepsilon,
\end{equation}
which corresponds to a ``shell'' of some thickness $\varepsilon$. As we will see, this choice leads to some nice properties consistent with other known results in the literature. This choice also allows us to reconcile with the result in \cite{2112.15418}, given in Eq.(\ref{e2}) and Eq.(\ref{e3}).

Indeed, if the total thermodynamic force is $f:= {\partial E}/{\partial r_+}= F_{\text{therm}}-1/4$, this suggests that a \emph{horizon pressure} can be defined via
\begin{equation}\label{pressure}
-\frac{1}{4}= - P \frac{\partial V}{\partial r_+} \Longrightarrow P\equiv P|_{r_+} = \frac{1}{32\pi r_+ \varepsilon}.
\end{equation}
For a Schwarzschild black hole, the first law with pressure term now leads to
\begin{flalign}
\d E &= T \d S - P \d V \\
&= \left(\frac{1}{4\pi r_+}\right) 2\pi r_+ \d r_+ - \left(\frac{1}{32\pi r_+ \varepsilon} \right)8\pi r_+ \varepsilon \d r_+ \\
&= \frac{1}{2}\d r_+ - \frac{1}{4} \d r_+=\frac{1}{4}\d r_+=\frac{1}{2}\d M.
\end{flalign}
We note that $\varepsilon$ drops out in the first law; though it might have a nice physical interpretation which we will return to in the next section. For now, we note from the previous calculation that the thermodynamic mass is half of the ADM mass: $E=M/2$.
The Smarr relation is readily verified to be 
\begin{equation}
E= 2TS - 2PV.
\end{equation}
We can also check that $PV=M/4$, thus it follows that 
\begin{equation}
2TS = E + 2PV = 2E,
\end{equation}
or simply $E=TS$. Therefore the result of \cite{2112.15418} (despite its heuristic derivation) can be recovered, once we realized that $E\neq M$ but instead $E = M/2$ as the result of the maximum force induced pressure term. 

One seemingly peculiar property is that while $V=0$ in the limit $\varepsilon \to 0$, $P \to \infty$ in the same limit, though their product is a constant. What happens in the $\varepsilon \to 0$ limit? In the following we will argue that $\varepsilon \to 0$ is equivalent to $\hbar \to 0$, thus we are requiring that in the classical limit black holes obey the original laws found by Bardeen, Carter and Hawking \cite{BCH}. For now, note that in the usual thermodynamics of a box of gas, the $PV$ term describes by the Boyle's law also satisfies $P \propto V^{-1}$, and so $P$ diverges in the $V \to 0$ limit. However, when $V=0$ there is \emph{no} box and so the system no longer exists and $E \equiv 0$. In our case, in the classical limit $V = 0$ and the $PV$ term is identically zero for the same reason. However, by virtue of the area law of black hole entropy, the $TS$ term remains, unlike the box of gas, and we recover the usual results: $E=M=2TS$. Note that the limit for the Smarr relation $\varepsilon \to 0$ is smooth although the limit of the product $PV$ is not. Even if one is ignorant of the presence of the pressure term, the thermodynamics is still effectively described by $E=TS$ or equivalently $M=2TS$, which would be the same as if there is indeed no pressure. If indeed $\varepsilon \to 0$ is equivalent to $\hbar \to 0$, then this is actually not as surprising as it sounds -- for the Smarr relation to hold classically (if one treats $M$ as classical), $\hbar$ must cancel out in the expression. That is to say, the value of $M$ must be independent of $\varepsilon$. The only difference is that in the case $\varepsilon \neq 0$, the ADM mass $M$ is distinct from the thermodynamic mass $E$.

Another remark is in order: note that $P>0$. In the literature, one finds various proposals that black hole (or horizonless compact object) interiors are filled with some kind of negative pressure fluid or other fields, e.g., \cite{1501.03806,1805.11667,1905.06799,2010.13225,2109.10017,2301.09712,2005.13260}. It should be emphasized that the value $\tilde{F}=-1/4$ is only for the horizon; so the interior can in principle still have a core with a negative pressure, which would be similar to the model in \cite{0506111}. In \cite{2108.06824}, one finds a different approach that also yields an effective pressure -- which can be either positive or negative -- from quantum gravitational correction. A pressure can also arise from regularizing the black hole singularity \cite{2304.05421}.

We also remark on another natural guess for the volume, namely $V=(4/3)\pi r_+^3$ the ``thermodynamic volume'' \cite{2010.13225,0701002}. This would lead to
$P=(16\pi r_+^2)^{-1}$.
However, the Smarr relation would lead to $E=2M/3$, which cannot be reconciled with \cite{2112.15418}. Note, however, that in \cite{2010.13225,0701002} the pressure is obtained from the stress-energy tensor, which vanishes for Schwarzschild, in which case $E=M$, which is quite unlike our situation.

\section{Interpreting the Schwarzschild Pressure}\label{IV}

We have come to the most speculative part of this work: an attempt to give a physical interpretation for the horizon pressure in the simplest case of Schwarzschild spacetime. This part is quite independent of the preceding sections in the sense that even if this interpretation turns out to be wrong it does not invalidate the prior results. 

As we have seen, the physical picture we have in mind is of a thin shell around the black hole horizon, which is similar to the membrane paradigm.
For the Schwarzschild case, we note that if we apply the 2-dimensional surface tension following the membrane paradigm \cite{D,PT,TPM},
\begin{equation}
\sigma := \frac{\kappa}{8\pi} = \frac{1}{32\pi M},
\end{equation}
then the pressure is related to the surface tension by
\begin{equation}\label{P1}
P = \frac{1}{32 \pi r_+ \varepsilon} = \frac{ \sigma }{2\varepsilon}.
\end{equation}

The question is whether we can determine the scale $\varepsilon$. This would require at least another equation that relates $P$ with $\varepsilon$. To this end, inspired by the Laplace's law of pressure for spherical membranes\footnote{A different notion of pressure was previously prescribed to the ``principal eigenvalue'' of the stability operator of marginally outer trapped surfaces \cite{1309.6593}.}, let us conjecture that the horizon pressure is also related to the thickness of the membrane $\varepsilon$ via
\begin{equation}\label{P2}
P = \frac{K\tilde{\sigma}\varepsilon}{r_+},
\end{equation}
where $\tilde{\sigma}$ is some quantuty akin to a ``wall stress'' (wall tension divided by wall thickness for the actual Laplace's law), and $K$ is a proportional (dimensionless) constant\footnote{Here we treat the fluid membrane to be close to the black hole of radius $r_+$. The correction due to using proper radius will also affect the prefactor, which can be absorbed into $K$. However, the distinction between proper distance and areal radius is not important to obtain the characteristic \emph{scale} of the fluctuation; c.f. Footnote 4.}. 
On dimensional ground, $\tilde{\sigma}$ is ${\sigma}$ divided by a length scale $L$. All we are arguing with Eq.(\ref{P2}) is that the horizon pressure is proportional to the surface tension and inversely proportional to the radius. However, \emph{unlike} the usual Laplace's law where $L=\varepsilon$, this cannot be so for our case, for otherwise $P$ is entirely independent of $\varepsilon$, contradicting Eq.(\ref{pressure}). Thus we choose the only natural length scale left in the system: $L=\ell_\text{P}$, the Planck length. 

Equating Eq.(\ref{P1}) and Eq.(\ref{P2}) yields the characteristic thickness (we keep $\ell_\text{P}$ explicit here for clarity)
\begin{equation}
\varepsilon \sim \sqrt{r_+ \ell_\text{P}},
\end{equation}
the geometric mean of the Schwarzschild radius and the Planck length $\ell_\text{P}$. We can check that Eq.(\ref{P2}) has the behavior that it diverges in the limit $\hbar \to 0$, which is the same as Eq.(\ref{pressure}). Our chain of arguments, from the maximum force induced pressure to conjecturing that a Laplace-like law holds finally led us to the length scale $\varepsilon \sim \sqrt{r_+ \ell_\text{P}}$. Remarkably, this scale has appeared a few times in the literature.

Notably, this is the scale of the quantum fluctuation of the horizon position, as previously shown by Marolf in \cite{0312059}
by examining how perturbation in quantum degrees of freedom at the temperature $T \sim 1/r_+$ could affect the horizon position (improving on Sorkin's result \cite{9701056}; refer also to \cite{9807065}). See also \cite{9306069,9606106v2}. In \cite{2205.01799}, Zurek provided a random-walk argument for this result. In fact, the $\sqrt{M}$ behavior is a characteristic of dissipative phenomena typically seen in hydrodynamics \cite{2304.12349} (in fact one can argue that a hydrodynamic behavior emerged from coarse-graining of the quantum physics). It is also the decoherence scale of nested causal
diamonds\footnote{Indeed it has been argued that such a fluctuation exists not only for black hole horizons but also for causal diamonds, which may have observational consequences using interferometer arms, much like in the set-up for gravitational wave detection \cite{2205.01799,2012.05870,2205.02233,1902.08207,2209.07543,2305.11224}.}, each of which has $S$ degrees of freedom (identified as entanglement entropy) \cite{2304.12349,2108.04806}.
Note that the horizon fluctuation scale is much larger than the Planck length, though still small. For example, for a solar mass black hole whose Schwarzschild radius is about $10^3~\text{m}$, the horizon fluctuation scale is about $\varepsilon \sim 10^{-16}~\text{m}$, which is subatomic. This scale also appears in the work of  Anastopoulos and Savvidou \cite{1410.0788}, in which they showed that the horizon of a black hole in thermal equilibrium with its Hawking radiation (``black hole in a box'') is surrounded by a thin shell of size $O(\sqrt{M})$ where they argued the Einstein equations break down. They suggested that this result should hold even when the box is removed and the system evolves slowly out of equilibrium, i.e., for an evaporating black hole. In \cite{1505.07131}, Brustein and Medved also argued that spacetime ceases to be semi-classical at this scale away from the classical horizon. Incidentally, this scale is associated with the wavelength of thermal radiation from a ball close to forming a black hole; see Appendix A of \cite{2003.10429}.   

The shell is thus not a classical fixed surface, but rather is consistent with the outer boundary of a ``quantum horizon region'' (QHR), the fluctuation of which induces a pressure on the effective classical horizon. Indeed, since black holes display special properties at $x=\sqrt{8/9}$, it is more natural to suspect that $\varepsilon$ is a quantity that has $x$ dependence, which $\varepsilon \sim \sqrt{r_+ \ell_\text{P}}$ provides (though we have not yet generalized to the charged/rotating case here), but another seemingly plausible choice $\varepsilon=\ell_\text{P}$ does not.

To recap, when quantum effects are not considered (hence also no Hawking temperature), the first law of black hole thermodynamics is the usual one with no pressure term, and WMFC holds -- black holes satisfy $F_\text{therm} \leqslant 1/2$ with equality attained for the Schwarzschild case; and $F_\text{therm}=1/4$ characterizes the transition to near-extremality\footnote{Though this is itself mostly motivated by peculiar behaviors in the -- quantum -- particle creation phenomena. But we can define near-extremality using $f$ first, fully incorporating quantum effects, and then translates that back into the condition on $F_\text{therm}$.}. 

If quantum effects are included so that the horizon location is uncertain, then a pressure term is introduced into the first law, with an associated force $\tilde{F}=-1/4$ and the total force $f = F_\text{therm} + \tilde{F}$ satisfies SMFC $f \leqslant 1/4$, and near-extremality is characterized by $f=0$. The claim that horizon fluctuation can be treated thermodynamically has been previously discussed in \cite{0803.4489}. Intuitively, a radially fluctuating horizon corresponds to a varying boundary condition for the quantum fields in the black hole vicinity. This is analogous to the dynamical Casimir effect\footnote{DeWitt showed that moving boundaries induce particle creation \cite{dw}. The case of an oscillating sphere was studied in \cite{0105282}. Black hole horizons have quite different boundary conditions so the analogy is not perfect. Nevertheless, perhaps we can learn some lessons from non-black hole systems \cite{2304.05992}.}, which can cause additional particle production in addition to the standard thermal one. The fluctuation thus leads to a Casimir-like force or pressure.
Indeed, horizon fluctuation is expected to only affect the spectrum of the radiation, not its temperature \cite{1109.6080}; see also \cite{1005.0286,1008.5059,0802.0658}. 

We remark that the horizon pressure Eq.(\ref{pressure}) is still much smaller than the Planck pressure, but it is huge by ordinary standard -- in SI units, for a solar mass black hole we have $P \sim 10^{54}~\text{Pa}$, which is larger than the pressure in a neutron star core, at $O(10^{35})$ Pa. Like Hawking radiation we should expect the value of the pressure to be observer dependent, and it is not clear at this point if a freely falling observer can detect it. Nevertheless, this pressure term hints at the possibility that once quantum effects are taken into account, black hole horizon is not uneventful (though not a divergent energy density as in a firewall \cite{0907.1190,1207.3123,1304.6483}) and the pressure might back-react on the spacetime. This is consistent with, e.g., \cite{2004.04956}.

\section{Discussion, Questions, and Future Prospects}\label{V}

To conclude, the fact that $F_\text{therm}=\partial M/\partial r_+ = 1/4$ corresponds to black hole parameters at which the spacetime displays some special properties can be used to characterize near-extremality, when some properties of the Hawking emission changes. However, these special properties are quite different for Reissner-Nordstr\"om and Kerr spacetimes, so we must proceed with caution. Future studies should clarify if various properties that occur at the value $Q/M=\sqrt{8/9}$ or $a/M=\sqrt{8/9}$ are somehow related and might have deeper connections, or simply coincidences.

We have also seen that \emph{imposing} SMFC as a physical principle naturally implies that black hole horizon has a pressure, and the thermodynamic mass is half that of ADM mass, which resolves the factor of two mystery between SMFC and WMFC. Interestingly the Smarr relation is equivalent to $E=TS$, since $PV=M/4=E/2$, which makes black holes more similar to a typical thermodynamical system.  
There are of course many questions that still need to be addressed. At the same time the maximum force viewpoint of horizon thermodynamics may also lead to new avenues for future research.

Firstly, although for the Schwarzschild case our pressure can be interpreted as the result of quantum fluctuation of the horizon position at the scale $\sqrt{M}$, the more general cases of Reissner-Nordstr\"om and Kerr black holes still require further works. It may turn out that this interpretation is not the correct one, though it seems hopeful. Marolf's argument for the thickness of the quantum horizon region is $\varepsilon \sim (r_+/T)^{1/4}$, whereas the argument presented in Sec.(\ref{IV}), by equating Eq.(\ref{P1}) and Eq.(\ref{P2}), would give $\varepsilon \sim 1/\sqrt{T}$, thus quantitatively these would not agree when there is an inner horizon. They agree qualitatively in the sense that $\varepsilon$ is larger when the black holes carry a charge or angular momentum. One has to be careful about what happens in the extremal limit (though a divergent $\varepsilon$ is not inconsistent with the location of where a typical Hawking particle is emitted, which extends to infinity in the extremal limit \cite{2003.10429}). Marolf pointed out that in his argument, the divergence in the extremal limit signals a breakdown of the near-horizon approximation used. In our argument, the conjectured ``Laplace's law'' would require correction due to electrical charges in the Reissner-Nordstr\"om case and non-spherical surface in the Kerr case. Indeed, whether Eq.(\ref{P2}) can be derived more rigorously would be an important task for future research. Presently, we can only justify it \emph{a posteriori} by the fact that it gives a physically interesting scale that corresponds to horizon fluctuation. That is, we can turn the logic the other way around: if we accept horizon fluctuation can induce an effective pressure term, this leads in turn to Eq.(\ref{P2}).
Note that a fluctuating horizon could also give rise to echoes in gravitational waves \cite{2202.09111}, which is potentially testable in future observations. 

Let us also address the quantum nature of thermodynamic force that corresponds to this pressure. It is true that $\hbar$ is absent in the maximum force conjecture, since the physical dimension of a force is simply $c^4/G$. This does \emph{not} mean that a force cannot be caused by quantum effects. A good -- and relevant -- example is the Casimir force (per unit area $A$) for an electromagnetic field, which in 4-dimensions, reads
\begin{equation}
\frac{F^{4d}_\text{Casimir}}{A}=-\frac{\pi^2}{240 a^4}\hbar c,
\end{equation}
where $a$ is the separation between the parallel plates. One notices that $\hbar$ is present. However, area and distance can be expressed as multiples of Planck area and Planck length, and \emph{all} the powers of $\hbar$ will cancel out so that the physical dimension of the Casimir force is $[F^{4d}_\text{Casimir}]=c^4/G$.
Nevertheless no one would seriously claim that the Casimir force  is a classical phenomenon, as it is the result of the fluctuation of the quantum field in between and around the plates. Even if one takes the view that Casimir effect is just Van der Waals' \cite{1605.04143}, the latter is still not classical; it is a many-body quantum effect\footnote{In other words, the distinctions here being: any force is \emph{eo ipso} a classical entity; whereas the fact that Casimir force in particular only arises from quantum effects means it \emph{ipso facto} has a quantum nature. Whether a physical quantity itself is more fundamental than its cause is a philosophy problem best left for the relevant experts. See \cite{0110060} for a related debate on what counts as ``fundamental constants''.}. For similar reasons, quantum fluctuation of the horizon location can give rise to a force. In fact, horizon fluctuation is more akin to the dynamical Casimir effect. 

Interestingly, in 2-dimensions, in which the Casimir force between two parallel plates is strongest, it is (for electromagnetic field)
\begin{equation}
{F^{2d}_\text{Casimir}}=-\frac{\pi}{12 a^2}\hbar c.
\end{equation}
Clearly Casimir force becomes stronger if the distance between the plates decreases.
If we naively take $a \to \ell_{\text{P}}$, then in Planck units we see that $|F^{2d}_\text{Casimir}|=\pi/12 \sim 1/4$. This is most probably just a coincidence, and the result is not exact anyway due to higher order terms and possibly new physics at the Planck scale (such as effect from the minimal length \cite{0502142}). This is not to say that Casimir forces in general cannot ever exceed the maximum force bound --
the point is that we should ask whether the maximum force conjecture applies more widely in other contexts that involve \emph{quantum} forces beyond the black hole scenario, not just in classical systems. 

{It is worth mentioning that classical quantities free of $\hbar$ arising from quantum phenomena are nothing new. For example, the Maxwell-Boltzmann distribution function that describes the statistical thermodynamics of ideal gas is free of $\hbar$ since the powers of $\hbar$ in the density of states canceled with that in the partition function. In fact, one notes that $\hbar$ can be absent in a quantum context, and can be present in a classical context. One sees this in the quantum partition functions of both the Bose-Einstein and Fermi-Dirac statistics, which do not explicitly contain $\hbar$, but the classical partition function of a gas of $N$ identical classical particles in 3 dimensions is an integral that contains a factor of $\hbar^{3N}$ in its denominator, namely:
\begin{flalign}
Z=&\frac{1}{N!(2\pi\hbar)^{3N}} \int d^3q_1 d^3q_2 \cdots d^3q_N  \int d^3p_1 d^3p_2 \cdots d^3p_N  \\ \notag
&\times \exp\left[-\beta \sum_{i=1}^N H(q_i,p_i)\right]. 
\end{flalign}
In terms of the trace, the quantum partition function is $Z=\text{tr}(e^{-\beta\hat{H}})$, which has no $\hbar$ since the argument of the exponential function is of course dimensionless.
Passing to continuous integral for the classical form, however,
\begin{equation}
Z=\int \langle x,p \left|e^{-\beta\hat{H}}\right|x,p\rangle \frac{dx dp}{2\pi\hbar},
\end{equation}
in which $\hbar$ appears due to the minimum cell size in the phase space.
For another example, consider the ``classical'' massive Klein-Gordon equation 
\begin{equation}
\left(\Box + \frac{m^2c^2}{\hbar^2} \right)\phi(x)=0.
\end{equation}
Although it is sometimes argued that $\hbar$ is just there for dimensional reason in these classical contexts, the deeper reason is that our world is inherently quantum.
Since we are discussing thermodynamics and statistical mechanics, we also note that the SI units of temperature, the Kelvin, is defined in terms of the Boltzmann constant, which has units $\text{JK}^{-1}$, which in turn requires $\hbar$ to define, though temperature as we learned in high schools can be discussed without quantum mechanics. In fact, as we have previously seen, since Planck units are comprised of $G, c, \hbar$, quantities like length and mass are necessarily multiples of the Planck length and Planck mass, respectively. The point I wish to emphasize is this: the fact that the purported maximum force is $c^4/G$ in no way excludes the possibility that it has a quantum origin, and over-emphasizing that it is ``classical'' just because of an absent of $\hbar$ could be misleading.

Returning to the discussion on the maximum force,} let us also note that, in our analysis on the thermodynamic force, SMFC was imposed by adding a constant force $\tilde{F}=-1/4$ to the thermodynamic force. This is not the only possibility. We can contemplate other choices of $\tilde{F}$, e.g., a non-constant function $\tilde{F}$ that satisfies $\tilde{F}(x=0)=-1/4$ and $\tilde{F}(x=1)=1/4$. An explicit example would be to demand that $f\equiv 1/4$ always holds for black holes (though we will lose the nice correspondence between the violation of maximum force and inner horizon instability), i.e.,
\begin{equation}
\tilde{F}=\frac{1}{4}-F_\text{therm}=\frac{1}{4}-\frac{\sqrt{1-x^2}}{1+\sqrt{1-x^2}},
\end{equation}
which yields a pressure of the form
\begin{equation}\label{press}
P=-\frac{1}{8\pi r_+ \varepsilon}\left[\frac{1}{4}-\frac{\sqrt{1-x^2}}{1+\sqrt{1-x^2}}\right].
\end{equation}
This choice of $\tilde{F}$ therefore yields a negative pressure at the horizon for near-extremal black holes, i.e., for $x > \sqrt{8/9}$. Whether this or other choices of $\tilde{F}$ is better (e.g., lead to more interesting physics) remains to be seen.

The thermodynamic force formulation of black hole thermodynamics and the constraints from the maximum force conjecture could yield further insights into the properties of black hole spacetimes.
For example, can we use it to constrain the interior fluid models of black holes?
Furthermore, the maximum force conjecture has been argued to hold in some other modified gravity theories \cite{2006.07338,2111.00212,2201.10381} (though the values of the maximum force could differ). It would be interesting to check if the saturation of thermodynamic force $F_{\text{therm}}=F_\text{max}$ coincides with special properties such as effective negative temperature of the horizon for charged black holes, and ISCO coinciding with the ergosphere for the rotating case. If not, this may give a strong evidence for 4-dimensional general relativity being unique from the maximum force perspective, though for a different reason than those advocated in \cite{2205.06302}. Incidentally, even in GR, it would be interesting to check if in the Kerr-Newman case one can find any peculiarity when the black hole parameters satisfy $F_\text{therm}=1/4$.

In this work, it is also noted that the inner horizon instability seems to be reflected by the fact that its associated force has a magnitude of $|f| > 1/4$. 
This property is automatically satisfied once the shift forced $f$ that satisfies SMFC is defined and applied to the inner horizon, which lends credence to the definition.
It would be interesting to study how this changes in the dynamical cases \cite{0808.1709,2001.11156,2308.09225}.
As previously mentioned, instability was also established in the case of cosmic string with tension whose magnitude exceeds the conjectured bound -- the exterior spacetime would collapse onto the string.
How much can we say about the possible connection between maximum force conjecture and instability of the spacetime? 
Let us also note that it is not difficult to violate $|f| > 1/4$ even in general relativity. For example, consider the Taub-NUT spacetime, with horizon $r_+=m+\sqrt{m^2+n^2}$, where $n$ is the NUT charge and $m$ its mass. We have
\begin{equation}
F^{\text{Taub-NUT}}_{\text{therm}}=\frac{\partial m}{\partial r_+} = 1 + \frac{1}{2}\frac{n^2-r_+^2}{r_+^2},
\end{equation}
and consequently the shifted force can be computed to be
\begin{equation}
f^{\text{Taub-NUT}}=\frac{\partial m}{\partial r_+} - \frac{1}{4} = \frac{1}{4}\left[\frac{2n^2+(1+\sqrt{1+n^2})^2}{(1+\sqrt{1+n^2})^2}\right],
\end{equation}
which is clearly larger than $1/4$ for any $n \neq 0$. This is consistent with the previous result that Taub-NUT spacetime is unstable \cite{0602045v1}. Since Taub-NUT spacetime is infested with closed timelike curves, could the maximum force bound play the role of chronology protection agent \cite{agent}? Here we must be cautious that Taub-NUT is not asymptotically flat. It would be important to study in the future whether the shifted force $f$ should be defined by subtracting a different constant that depends on the asymptotic behavior, in a similar spirit to ``background subtraction''. In particular, it is interesting to investigate whether this notion can be generalized to asymptotically locally anti-de Sitter spacetimes, which are important for holography. If $f$ can be understood in general, and if $|f|>c$ for some $c$ does indicate some sort of instability, then this would provide us a relatively simple method to determine stability. 

Another important question is to understand whether the thermodynamic force can be given a more rigorous foundation. As it is, as noted in Footnote 2, $\partial M/\partial r_+$ is clearly coordinate dependent, so changing to another coordinate system with radial coordinate $R=R(r)$, say, then $\partial M/\partial R$ would not give the same value. One possible explanation is as follows: recall that in the well-known Buchdahl bound \cite{Buchdahl}, the statement $M \leqslant 4r/9$ for a static fluid sphere (a ``star'') of radius $r$ is also coordinate dependent, but an invariant version can be formulated by replacing $r$ with the physical (proper) radius of the star $r^*$, thereupon the bound becomes $M \lesssim 0.3404 r^*$. 
Therefore, the thermodynamic force can be interpreted invariantly in the same manner if one takes $2M$ to be the proper size of a black hole radius. Indeed, in the Schwarzschild spacetime, a ``proper distance coordinate'' $R^*$ can be defined as \cite{TPM}:
\begin{equation}
R^*:=2M+\sqrt{r(r-2M)}+\ln\left[\sqrt{\frac{r}{2M}-1}+\sqrt{\frac{r}{2M}}\right],
\end{equation} 
so that $d(R^*):=R^*-2M$ measures the proper radial distance away from the horizon. From the exterior perspective, it makes sense to take the black hole to have a proper size of radius $2M$. Then $R^*=r_+=2M$ for the horizon would give the same thermodynamic force value. It would be interesting to explore other ways to make the concept of thermodynamic force more rigorously defined. Related to the issue of coordinate dependence is whether we should define the thermodynamic force by taking derivative with respect to the Schwarzschild or the Schwarzschild diameter (it is a special case of choosing a coordinate $R=2r$). In \cite{2205.06302}, the maximum force is attributed to the energy of a black hole distributed along the diameter (note that this only makes sense from an exterior observer's point of view, who treats the interior blindly as an ordinary sphere instead of a dynamical spacetime). While intuitive, it is hard to make this rigorous. For example, what happens in the Kerr case when the horizon is not spherical? In some sense this is similar to the well-known Hoop conjecture, whose exact formulation (what the appropriate notion of ``mass'' and ``size'' should be) is highly nontrivial \cite{0912.4001}. 

On the other hand, the examples discussed in 
\cite{2102.01831,2207.02465} demonstrated that one should be careful about what kind of forces are relevant for the conjecture (see also the local vs. quasi-local notions in \cite{PhysRevD.104.068502}). Let us note that for both the cosmic string and the black hole inner horizon, the system under consideration is ``non-classical'' in the sense that they both violate the classical energy conditions (and for the latter, even exhibit negative effective temperature); even Taub-NUT spacetime contains negative energy between the horizon at $r_+$ and infinity \cite{0602045v1}. It would not be surprising that energy conditions and how matter fields are coupled to gravity will play a crucial role in the final understanding of the maximum force conjecture and its possible relation to spacetime instability, and perhaps to cosmic censorship and chronology protection.

Of course, it might be interesting to investigate whether the pressure term proposed above gives rise to any new effect in the context of maximum force applied to cosmological horizons. 

Despite the speculative nature in some parts of this work, my hope is that it will provide some new ways to think about black hole thermodynamics, and also demonstrates that taking thermodynamic pressure and its associated volume into account provides a novel perspective to look at the maximum force conjecture in the contexts of black holes. 

\begin{acknowledgments}
The author thanks Brett McInnes for useful suggestions. He also thanks various colleagues at RIKEN Interdisciplinary Theoretical and Mathematical Sciences Program (iTHEMS) and the Kobayashi-Maskawa Institute of Nagoya University for hospitality, as well as useful questions and feedback during his visit and sharing of the preliminary findings of this work. 
\end{acknowledgments}

\end{document}